\documentclass[osajnl,twocolumn,showpacs,superscriptaddress,10pt]{revtex4-1}

\usepackage{graphicx}
\usepackage{amsmath}
\usepackage{amssymb}
\usepackage{color}
\usepackage{units}
\usepackage[latin1]{inputenc}
\usepackage{verbatim}
\usepackage{float}
\usepackage[english]{babel}
\usepackage{booktabs}	
\usepackage{epstopdf}
\usepackage{pgf}
\usepackage{pgfplots}
\usepackage{placeins}

\begin{document}

\title{Cancellation of lateral displacement noise of 3-port gratings \\ for coupling light to cavities} 

\author{Melanie Meinders}
\affiliation{Institut f\"ur Laserphysik und Zentrum f\"ur Optische Quantentechnologien der Universit\"at Hamburg,\\%
Luruper Chaussee 149, 22761 Hamburg, Germany}
\affiliation{Institut f\"ur Gravitationsphysik der Leibniz Universit\"at Hannover and\\ %
Max-Planck-Institut f\"ur Gravitationsphysik (Albert-Einstein-Institut), Callinstra{\ss}e 38, 30167 Hannover, Germany}

\author{Stefanie Kroker}
\affiliation{Institute of Applied Physics, Abbe Center of Photonics, Friedrich-Schiller-Universit\"at Jena,\\Max-Wien-Platz 1, 07743 Jena, Germany}%

\author{Amrit Pal Singh}
\affiliation{Institut f\"ur Laserphysik und Zentrum f\"ur Optische Quantentechnologien der Universit\"at Hamburg,\\%
Luruper Chaussee 149, 22761 Hamburg, Germany}
\affiliation{Institut f\"ur Gravitationsphysik der Leibniz Universit\"at Hannover and\\ %
Max-Planck-Institut f\"ur Gravitationsphysik (Albert-Einstein-Institut), Callinstra{\ss}e 38, 30167 Hannover, Germany}%

\author{E.-Bernhard Kley}
\author{Andreas T\"unnermann}
\affiliation{Institute of Applied Physics, Abbe Center of Photonics, Friedrich-Schiller-Universit\"at Jena,\\Max-Wien-Platz 1, 07743 Jena, Germany}%

\author{Karsten Danzmann}
\affiliation{Institut f\"ur Gravitationsphysik der Leibniz Universit\"at Hannover and\\ %
Max-Planck-Institut f\"ur Gravitationsphysik (Albert-Einstein-Institut), Callinstra{\ss}e 38, 30167 Hannover, Germany}%

\author{Roman Schnabel}
\email[corresponding author:\\ ]{roman.schnabel@physnet.uni-hamburg.de}
\affiliation{Institut f\"ur Laserphysik und Zentrum f\"ur Optische Quantentechnologien der Universit\"at Hamburg,\\%
Luruper Chaussee 149, 22761 Hamburg, Germany}
\affiliation{Institut f\"ur Gravitationsphysik der Leibniz Universit\"at Hannover and\\ %
Max-Planck-Institut f\"ur Gravitationsphysik (Albert-Einstein-Institut), Callinstra{\ss}e 38, 30167 Hannover, Germany}%

\begin{abstract} 
Reflection gratings enable light coupling to optical cavities without transmission through substrates. Gratings that have three ports and are mounted in second-order Littrow configuration even allow the coupling to high-finesse cavities using low diffraction efficiencies. In contrast to conventional transmissive cavity couplers, however, the phase of the diffracted light depends on the lateral position of the grating, which introduces an additional noise coupling. Here we experimentally demonstrate that this kind of noise cancels out once both diffracted output ports of the grating are combined. We achieve the same signal-to-shot-noise ratio as for a conventional coupler. From this perspective, 3-port grating couplers in second-order Littrow configuration remain a valuable approach to reducing optical absorption of cavity coupler substrates in future gravitational wave detectors.
\end{abstract} 

\maketitle

\begin{figure}[b]
\center
\includegraphics[width=\columnwidth]{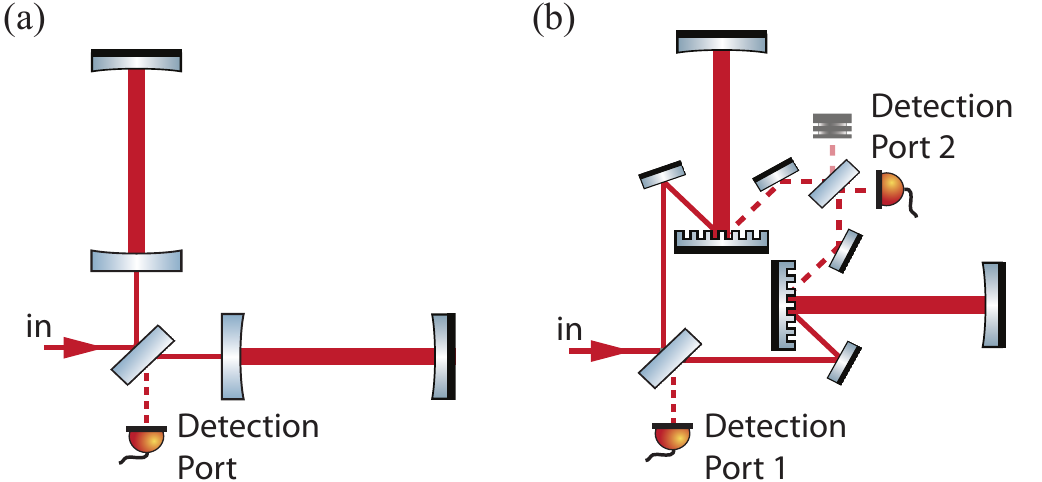}
\caption{ \textbf{Michelson interferometers with arm resonators. }
(a) Michelson-type laser interferometer with conventional, transmissively coupled Fabry-Perot arm cavities. 
(b) Equivalent interferometer topology with arm cavities that are diffractively coupled in second-order Littrow configuration. Here, the phase signal that is acquired in an arm cavity couples out via two ports of the grating. We show that summing up the photo-electric currents as shown not only recovers the full signal-to-shot-noise ratio but also cancels out phase noise due to lateral grating motions.}
\label{fig1:ifos}
\end{figure}

Coupling light to an optical cavity is usually realized via partially transmissive mirrors. 3-port reflection gratings represent a practical alternative, which does not require light transmission through substrate material \cite{Sun,Bunkowski2004,Bunkowski2005,Clausnitzer2005,gratingbs, higheff1,Bunkowski2006,Burmeister2010,Barr2011,Britzger2011,Britzger2012}. This avoids light absorption in substrates and consequently even allows opaque substrates. Due to this advantage reflection gratings were considered as optical components in gravitational-wave detectors already 20 years ago \cite{Drever1995}. 
Fig.~\ref{fig1:ifos} shows an example how (a) conventional couplers to arm cavities in a Michelson interferometer can be replaced with (b) weak diffraction efficiency gratings in second-order Littrow configuration \cite{Bunkowski2004,Bunkowski2005,Clausnitzer2005,Bunkowski2006,Burmeister2010,Barr2011,Britzger2011,Britzger2012}. 
Unfortunately it turned out that lateral grating displacements that are parallel to its surface and perpendicular to the grating's grooves cause phase shifts on the diffracted light field and thus introduce a new noise coupling \cite{Wise2005,Barr2011}. 
This contrasts a conventional mirror that is displaced parallel to its surface. It thus seems that gratings need to have a better isolation from environmental disturbances than conventional mirrors \cite{Freise2007}.

In this work we experimentally show that the unwanted lateral-displacement-to-phase coupling in grating interferometers can be canceled by an appropriate combination of detection ports. We realize a simplified setup in which a phase-modulated light beam is sent under normal incidence to a grating, whose position is continuously changed perpendicular to its grooves and to the optical axis of the incident light. We detect the phase modulations in the $\pm$ first diffraction orders and demonstrate that  summing up the photo-electric currents not only retains the full signal-to-shot-noise ratio but also cancels out phase noise due to the lateral grating motion.

\begin{figure}[b!]
 \center
\includegraphics[width=\columnwidth]{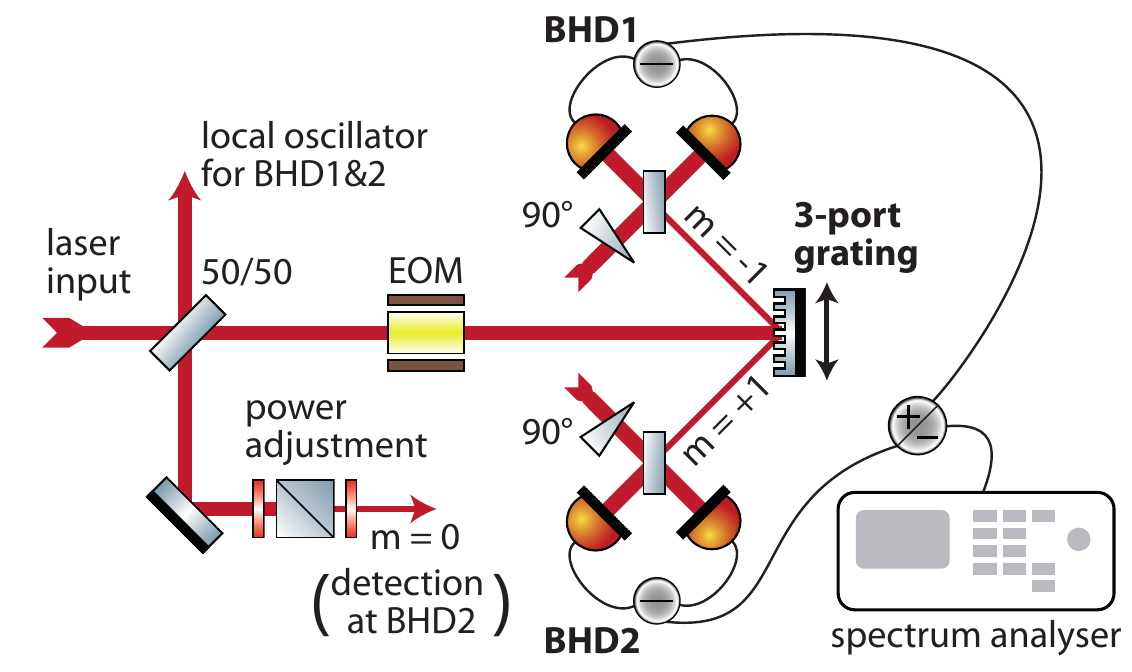}
\caption{ \textbf{Experimental setup. }We performed a phase measurement of the $m=\pm1$ diffraction orders of a 3-port grating  with two balanced homodyne detectors (BHD1\&2). The sum and difference of both signals were acquired with a spectrum analyzer. The position of the grating was modulated in the lateral direction, using a piezo actuator. An electro-optic modulator (EOM) periodically shifted the phase of the incident light, serving as a reference `gravitational-wave'-signal. Part of the zero order diffracted light was picked off at a 50/50 beam splitter and (optionally) detected with BHD2 to confirm the lateral displacement of the grating.}
\label{fig2:setup}
\end{figure}
\begin{figure}[t]
 \center
\includegraphics[width=\columnwidth]{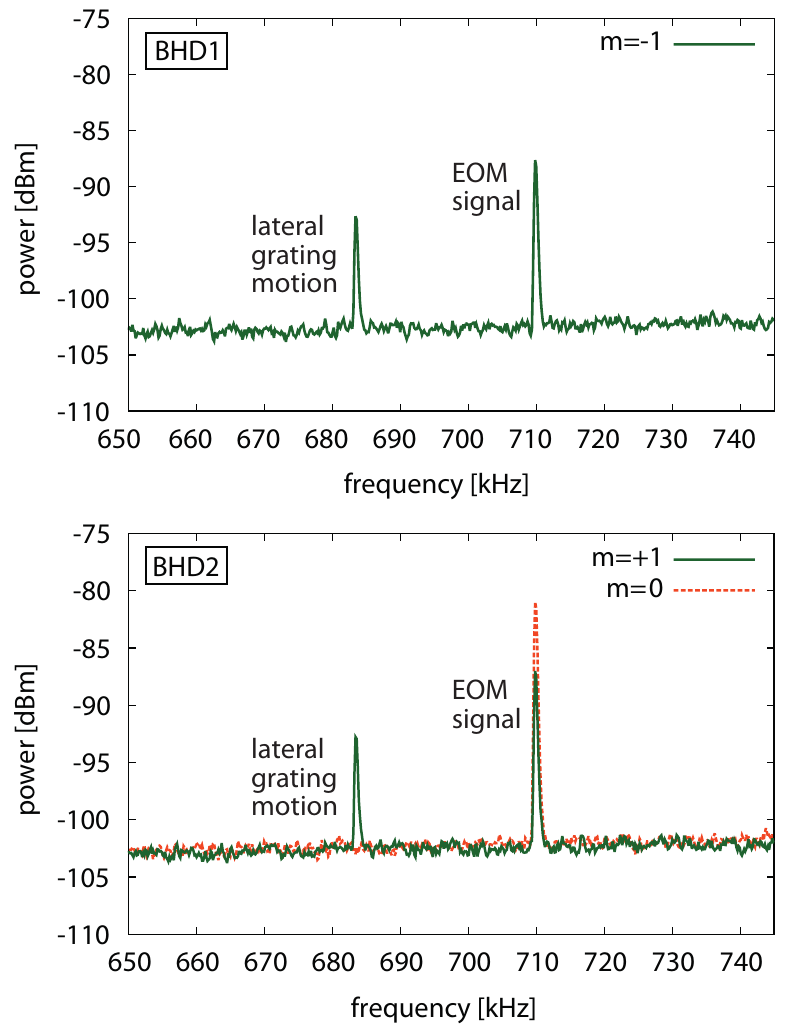}
\caption{ \textbf{Individual diffraction orders. }
Shown are the power spectra of the $\pm 1^{\rm{st}}$ diffraction orders (solid green) and the $0^{\rm{th}}$ order (dashed orange) as measured by BHD1\&2, respectively. During all these measurements the grating was laterally displaced at a frequency of about 683\,kHz  and a phase modulation at 710\,kHz was imprinted on the input light by an electro-optic modulator (EOM). The disturbance from the grating displacement is clearly visible in the measurements on the $\pm 1^{\rm{st}}$ diffraction orders. As expected it is not present in the $0^{\rm{th}}$ order.  (resolution bandwidth (RBW): 300\,Hz; video bandwidth (VBW): 10\,Hz; averaged 16 times). 
}
\label{fig3:individual}
\end{figure}

The schematic of the experimental setup is depicted in Fig.~\ref{fig2:setup}. Our 3-port diffraction grating had a size of 10\,mm\,x\,10\,mm and was realized on a quartz substrate with a size of 1''\,x\,1''\,x\,0.25''. 
The grating was realized using electron beam lithography to define the grating pattern into a resist layer. Afterwards this pattern was transferred by an inductively-coupled plasma etching process (ICP) into a chromium and subsequently into the uppermost silica layer. 
The first-order diffraction efficiencies at normal incidence were measured to be about 4.8\% at 1064\,nm. The normal incidence light beam had a power of 7\,mW, which resulted in 0.34\,mW in the $m=\pm1$ diffraction orders. The input light was modulated at a frequency of 710\,kHz by an electro-optic modulator (EOM), which served as a scientific phase signal corresponding to a `gravitational-wave'-signal. Two balanced homodyne detectors (BHD1\&2) measured the phase quadrature amplitudes of the first-order diffracted outputs. Each BHD was stabilized to this quadrature by using the difference of its photo diodes' DC voltage as an error signal and by feeding back to a piezo-actuated mirror in the path of the local oscillator beam (DC-lock). The unity gain frequencies for these locks were far below the frequencies of the injected signals and the signs for the quadratures were chosen in such a way that the radio-frequency EOM signal added when combining the BHD photo currents. The BHDs' local oscillator powers were 10\,mW. 
The individual BHD photo currents, as well as their sum (or difference) were analyzed with a spectrum analyzer. The grating was piezo actuated and modulated in the lateral direction at a frequency of about 683\,kHz in order to produce the disturbance phase signal under investigation here. 
The frequency of about 683\,kHz was selected because the piezo-actuated grating including its mount showed a pure lateral displacement in this case; any motion in direction of the incident laser beam was not visible, i.e.~was far below the shot-noise level of our setup. 

Fig.~\ref{fig3:individual} shows the individual measurements of the $m=\pm1$ diffraction orders at the respective detectors BHD1\&2 (solid green lines).
The disturbances from the lateral grating displacement and the scientific EOM signals are clearly visible. 
The broadband noise floor in our measurements was given by optical shot-noise. The transfer functions of the BHDs were almost identical and produced a marginal slope on the otherwise white shot noise.
To confirm that the grating signal was solely due to lateral displacement, we also analyzed the zero order diffraction. The result is shown in Fig.~\ref{fig3:individual} (bottom, dashed orange line). This measurement was done by picking off part of the zero order diffracted beam at a 50/50 beam splitter in the input path, adjusting the power to the same level as in the $m=\pm1$ diffraction orders and detecting it with BHD2 (via flip mirror). The measurement confirms that the grating signal vanished in zero order diffraction while the reference EOM-signal was clearly visible.
As expected, the measured EOM-signal was in fact about 6\,dB larger than in the first-order diffracted outputs, because the zero order light passed the EOM twice. 
The modulations during the passages were approximately in phase because the wavelength of the kHz-signal was much larger than the optical path between the grating and the EOM. The modulation amplitude was therefore doubled and in the depicted power spectrum thus resulted in a factor of 4 or 6\,dB accordingly. 

\begin{figure}
\center
\includegraphics[width=\columnwidth]{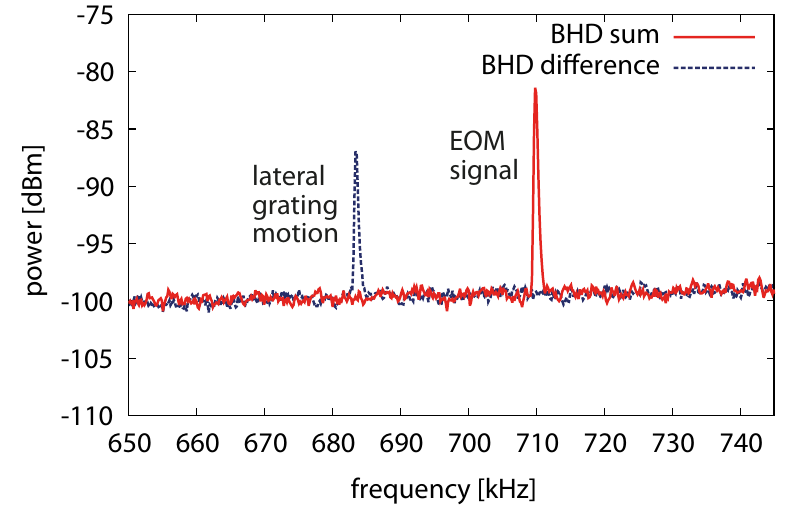}
\caption{ \textbf{Combination of the {\bf{$\pm 1^{\rm{st}}$}} diffraction orders. } 
Shown are the spectra of the sum (red) and difference (blue) of the $m=\pm1$ diffraction orders, measured by BHD1\&2. The disturbance signal that was produced by lateral displacement of the grating clearly vanished in the \emph{sum}, corresponding to a cancellation of the lateral-displacement-to-phase coupling of the grating. As expected, the phase modulation imprinted on the incident beam by the EOM increased by 6\,dB in this case. On the contrary, the difference photo current shows an increase of the disturbance signal and a completely vanished scientific phase signal. As expected for shot-noise limited measurements, the shot-noise level increased by 3\,dB in either case.}
\label{fig4:results}
\end{figure}

In Fig.~\ref{fig4:results} the sum (red) and difference (blue) of the two homodyne measurements are depicted. Since the optical shot-noise measured at the two detectors is uncorrelated, it adds up in variance and the noise floor was increased by 3\,dB in these measurements. As expected, the phase modulation from the lateral displacement of the grating cancels in the sum of the outputs while the amplitude of the scientific signal from the EOM adds up. 

In conclusion, using a simplified setup, we have experimentally demonstrated the cancellation of interferometric disturbance signals that result from lateral displacements of gratings. We considered the case in which a 3-port reflection grating mounted in second order Littrow configuration served as a cavity coupler, which corresponded to an arrangement that is potentially interesting for future gravitational wave detectors (Fig.~\ref{fig1:ifos} (b)). When the output ports were combined in a way that optimized the signal-to-shot-noise, the disturbance signals from lateral grating displacements canceled completely with respect to the shot noise of our setup. 
Our result suggests that phase noise from lateral displacement of 3-port gratings can be canceled to a high degree and that no additional demands on the suspensions are required,  as long as both diffracted outputs are being detected.\\


\begin{acknowledgments}

This work was supported by the Deutsche Forschungsgemeinschaft (Sonderforschungsbereich Transregio 7, project C3) and by the International Max Planck Research School for Gravitational Wave Astronomy (IMPRS-GW).

\end{acknowledgments}


\end{document}